\documentclass[useAMS,usenatbib]{mnras}

\usepackage{newtxtext,newtxmath}
\usepackage[T1]{fontenc}

\usepackage{graphicx}
\usepackage{grffile}
\usepackage{caption}
\usepackage{subcaption}

\newcommand{\kmsmpc}{\kms\;{\rm Mpc}^{-1}}

\newcommand{\kms}{{\rm km}\,{\rm s}^{-1}}
\newcommand{\cms}{{\rm cm}^{-2}}
\newcommand{\cmc}{{\rm cm}^{-3}}

\newcommand{\Zsolar}{\;{\rm Z}_{\odot}}
\newcommand{\msolar}{{\rm M}_{\odot}}

\newcommand{\photonssarcmin}{{\rm photons\, s}^{-1}\, {\rm arcmin}^{-2}}
\newcommand{\lxunit}{{\rm erg\; s}^{-1}}

\newcommand{\gad}{{\sc Gadget-3}}
\newcommand{\OVI}{\hbox{O\,{\sc vi}}}

\newcommand{\HI}{{\hbox{H\,{\sc i}}}}

\newcommand{\MgII}{{\hbox{Mg\,{\sc ii}}}}

\newcommand{\nh}{{n_{\rm H}}}
\newcommand{\LXextended}{L_{X,>10{\rm kpc}}}
\newcommand{\kco}{{\kappa}}

\begin{document}

\title[X-ray CGM of edge-on discs and spheroids]{Predictions for the X-ray circumgalactic medium of edge-on discs and spheroids}

\author[A. Nica et al.]{
\parbox[t]{\textwidth}{\vspace{-1cm}
Anna Nica$^{1}$\thanks{anna.nica@colorado.edu}, Benjamin D. Oppenheimer$^{1}$, Robert A. Crain$^{2}$, \'{A}kos Bogd\'{a}n$^{3}$, Jonathan J. Davies$^{4}$, William R. Forman$^{3}$, Ralph P. Kraft$^{3}$, John A. ZuHone$^{3}$
}\\\\
$^1$CASA, Department of Astrophysical and Planetary Sciences, University of Colorado, 389 UCB, Boulder, CO 80309, USA\\
$^2$Astrophysics Research Institute, Liverpool John Moores University, 146 Brownlow Hill, Liverpool, L3 5RF, UK\\
$^3$Harvard-Smithsonian Center for Astrophysics, 60 Garden St., Cambridge, MA 02138, USA\\
$^4$Department of Physics and Astronomy, University College London, Gower Street, London WC1E 6BT, UK
}
\maketitle

\pubyear{2021}

\maketitle

\label{firstpage}


\begin{abstract}

We investigate how the X-ray circumgalactic medium (CGM) of present-day galaxies depends on galaxy morphology and azimuthal angle using mock observations generated from the EAGLE cosmological hydrodynamic simulation.  By creating mock stacks of {\it eROSITA}-observed galaxies oriented to be edge-on, we make several observationally-testable predictions for galaxies in the stellar mass range $M_\star=10^{10.7-11.2}\;$M$_{\odot}$.  The soft X-ray CGM of disc galaxies is between 60 and 100\% brighter along the semi-major axis compared to the semi-minor axis, between 10-30 kpc.  This azimuthal dependence is a consequence of the hot ($T>10^6$ K) CGM being non-spherical: specifically it is flattened along the minor axis such that denser and more luminous gas resides in the disc plane and co-rotates with the galaxy.  Outflows enrich and heat the CGM preferentially perpendicular to the disc, but we do not find an  observationally-detectable signature along the semi-minor axis.  Spheroidal galaxies have hotter CGMs than disc galaxies related to spheroids residing at higher halos masses, which may be measurable through hardness ratios spanning the $0.2-1.5$ keV band.  While spheroids appear to have brighter CGMs than discs for the selected fixed $M_\star$ bin, this owes to spheroids having higher stellar and halo masses within that $M_\star$ bin, and obscures the fact that both simulated populations have similar total CGM luminosities at the exact same $M_\star$.  Discs have brighter emission inside 20 kpc and more steeply declining profiles with radius than spheroids.  We predict that the {\it eROSITA} 4-year all-sky survey should detect many of the signatures we predict here, although targeted follow-up observations of highly inclined nearby discs after the survey may be necessary to observe some of our azimuthally-dependent predictions.  

\end{abstract}


\section{Introduction} \label{sec:intro}

The circumgalactic medium (CGM) is the gaseous baryonic component surrounding a galaxy.  Over cosmic times, these CGMs are thought to participate in the assembly and evolution of galaxies.  The properties of the CGM have been observed to exhibit azimuthal dependence around highly inclined galaxies in UV absorption line surveys.  $\MgII$ shows azimuthal dependence within $\sim 50$ kpc such that absorption is stronger along the polar, semi-minor axis, and along the equatorial, semi-major axis \citep{bordoloi11,kacprzak12,bouche12,lan14, ho17,zabl19,martin19}.  It is less clear if high excitation UV ions like $\OVI$ display a similar azimuthal dependence \citep{kacprzak15,kacprzak19}, though \citet{beckett21} finds $\OVI$ enhanced along the semi-minor axis.  The interpretation often is that UV absorption along the polar direction is preferentially tracing superwind-driven outflows, while gas along the equatorial direction is tracing inflows \citep[e.g.][]{shen13, mitchell20a, peroux20}, and may thus present a revealing view of the baryon cycle in action.  

Soft X-ray emission around disc-like galaxies also exhibits azimuthal dependence; however the number of detected galaxies and the detection distance from the galaxy are limited by the capability of existing X-ray telescopes.  Emission from edge-on galaxies usually does not extend far beyond the optical extent of the disc, this being limited to the disc-halo interface \citep{li08,hodgeskluck13,li17,hodgeskluck18} and may more appropriately be considered an extended interstellar medium (ISM).  While X-ray emission is observed extended above and below the discs of starbursting galaxies \citep[e.g.][]{strickland04,hodgeskluck20} indicating clear signatures of outflows \citep{strickland09}, these objects are rare outliers that are among the most luminous extended X-ray objects associated with disc galaxies.  

Extended X-ray emission from gaseous haloes around typical galaxies are a general prediction of cosmological hydrodynamical simulations \citep{toft02,rasmussen09,crain10,crain13,kelly21}; however, {\it Chandra} and {\it XMM-Newton} possess the sensitivity to detect emission associated with only a handful of the most massive nearby late-type galaxies \citep{anderson11,dai12,bogdan13a,bogdan13b,anderson16,bogdan17,li17,das19}.  Newer simulations, including EAGLE \citep{schaye15,crain15,mcalpine16} and IllustrisTNG \citep[][]{pillepich18,nelson18a}, have been tuned to fit some of the observed properties of galaxies.  X-ray emission predictions, for both EAGLE \citep{davies19} and IllustrisTNG \citep{truong20}, have been generated from these simulations and compared to observations of emission around galaxies, but these mainly focus on central regions and not extended emission for typical galaxies.

The {\it eROSITA} instrument on the Spectrum-Roentgen-Gamma mission \citep{erosita}, launched in July 2019, opens new possibilities for the detection of extended, soft X-ray emission ($\leq 2.0$ keV) associated with typical galaxies.  The on-board detectors are operating within mission expectations for soft X-ray energies \citep{predehl21}, which means that the grasp (the produce of the collecting area and the field of view) of {\it eROSITA} should achieve the expected signal in its primary 4-year all-sky survey comprising eight 6-month scans (the eRASS:8 survey).  {\it eROSITA} is predicted to detect $>10^5$ clusters and groups \citep{pillepich18c}, and it will also collect photons from the CGM of galaxies, but stacking methods are required to detect the emission as was done for the {\it ROSAT} all-sky survey \citep{anderson15}.  

The $30\times$ greater sensitivity of {\it eROSITA} compared to {\it ROSAT} all but assures the detection of extended X-ray emission associated with galaxies. The superior spatial resolution of {\it eROSITA} offers a better opportunity to separate CGM emission from galactic sources including X-ray binaries, the hot ISM, and potential active galactic nuclei (AGN).  \citet[][hereafter O20]{oppenheimer20b} predicted that {\it eROSITA} should be able to detect soft X-ray emission in the CGM around typical galaxies in the nearby Universe at $z\la 0.01$.  The aim of O20 was to demonstrate that {\it eROSITA} could resolve diffuse X-ray haloes around galaxies with masses as low as $M_\star=10^{10.2}\,\msolar$.  Given the prediction of both the EAGLE and IllustrisTNG simulations that more highly star-forming galaxies at fixed halo mass have denser CGMs \citep{davies20}, O20 predicted that {\it eROSITA} emission should grow stronger with higher star formation rates (SFRs) at fixed stellar mass.  

Here we continue the investigation of mock {\it eROSITA} observations by using the EAGLE simulation to forward model observations of galaxies rotated completely inclined to appear edge-on.  We define galaxy subsamples using kinematic morphology, rotating them all to be inclined, while concentrating on more massive galaxies with $M_{\star}=10^{10.7}-10^{11.2}\,\msolar$.  The purposes of our morphology-centred investigation is two-fold: 1) to determine if there is azimuthal dependence in simulated X-ray stacks of disc-like galaxies, and 2) to establish whether the emission depends on morphology.  The second purpose does not necessarily rely on galaxy orientation, and provides an exploration of X-ray emission as a function of morphology.  

We focus on mock {\it eROSITA} observations, because the eRASS surveys provide all-sky coverage with consistent quality and data reduction.  This contrasts with deep {\it XMM} and {\it Chandra} observations focused on individual galaxies that often fall in the category of rare outliers.  By choosing stacks of $\sim$100 galaxies within morphologically-similar samples, we develop an experiment to survey the appearance of typical galactic X-ray CGM profiles.  We create {\it idealised} samples by placing the galaxies at $z=0.005$, a distance of 22 Mpc, but in practice it will be necessary to use galaxies that either are at higher redshift, are at higher ecliptic latitudes where eRASS exposure times are longer, and/or include high inclinations as opposed to perfectly edge-on discs.  

The paper is arranged as follows. Section \ref{sec:methods} introduces the EAGLE simulation, defines the galaxy samples, and details the {\it eROSITA} forward modelling pipeline.  The main results are presented in \S\ref{sec:results}, first concentrating on the mock observational results in \S\ref{sec:obsres} and the physical interpretation within the EAGLE simulations in \S\ref{sec:physres}.  \S\ref{sec:discuss} discusses some of the results relating to X-ray CGM properties as a function of morphology, the significance of emission along the equatorial direction, existing observations of X-ray emission around inclined galaxies, and a discussion of the similar work by \citet{truong21} focusing on the IllustrisTNG simulation.  We summarise in \S\ref{sec:summary}.


\section{Methods} \label{sec:methods}

\subsection{EAGLE simulations} 

Our analysis uses the ``Reference'' EAGLE cosmological simulation \citep{schaye15,crain15}.   This $100^3$ comoving Mpc$^3$ run, referred to as Ref-L100N1504 uses initial conditions with $1504^3$ collisionless dark matter (DM) and an initially equal number of smooth particle hydrodynamic (SPH) particles starting at $z=127$.  The  \citet{planck13} cosmogony is used ($\Omega_{\rm m}=0.307$, $\Omega_{\Lambda}=0.693$, $\Omega_{\rm b}=0.04825$, $H_0= 67.77$ $\kmsmpc$). The code used is a significantly modified version of the N-body/Hydrodynamical code \gad~last described in \citet{springel05}.  The SPH implementation uses a pressure-entropy-based formulation \citep{hopkins13} and a series of additional modifications referred to as {\sc ANARCHY}, the influence of which are explored by \citet{schaller15}.  

The EAGLE code applies a number of subgrid physics modules, including radiative cooling \citep{wiersma09a}, star formation \citep{schaye08}, stellar evolution and metal enrichment \citep{wiersma09b}, super-massive black hole (SMBH) formation and accretion \citep{booth09,schaye15,rosas15}, stellar feedback \citep{dallavecchia12}, and SMBH feedback \citep{booth09}.  Thermal prescriptions, where the imparted feedback energy heats local SPH particles, are applied for both stellar and SMBH feedback.  \citet{crain15} describes how the calibration of these feedback schemes credibly reproduces the galactic stellar mass function and galaxy sizes.  

The EAGLE simulation has a mass resolution for DM particles of $9.7\times 10^6\ \msolar$ and for SPH particles of $1.8\times 10^6\ \msolar$.  This resolution has a Plummer-equivalent softening length of 700 proper pc at $z<2.8$, and 2.66 comoving kpc at $z>2.8$.  The inter-particle SPH separation is $3.8\times (\nh/(10^{-3} \cmc))^{-1/3}$ kpc.  

EAGLE haloes are identified via a two-step process, starting with a friends-of-friends algorithm linking DM particles within a length of 0.2 the mean inter-particle separation, and linking associated gas and star particles to the nearest DM particle.  The SUBFIND algorithm \citep{springel01, dolag09} then identifies bound substructures within the haloes, and the halo mass is characterised by the spherical overdensity mass ($M_{200}$) centred on the halo's most bound particle.

\subsection{Galaxy subsamples} \label{sec:samples}

To generate observationally reproducible samples, we select simulated central galaxies based on stellar mass and kinematically-defined morphology that has been shown to accurately correspond to direct measures of morphology \citep{thob19}.  We focus on the mass range of $M_\star = 10^{10.7-11.2}\,\msolar$, which is a bin of width $0.5$ dex that O20 labeled as the ``High-mass'' sample.  EAGLE has 498 galaxies in this mass range within its $10^6$ Mpc$^3$ volume.  Haloes with $M_{200}>10^{13.3}\,\msolar$, which are often considered groups, are excluded because we expect the CGM of such galaxies to be individually detectable with {\it eROSITA}.  

The resulting sample has 429 central galaxies in the given mass range for EAGLE, which is divided into quartile subsamples, each of 107 galaxies, using their kinematic morphologies.  We use the definition of $\kco$, which is the fraction of stellar kinetic energy invested in co-rotation\footnote{The variable $\kappa_{\rm co}$ is often used in other publications in place of $\kco$.}, to define our samples listed in Table \ref{tab:samples}.  \citet{correa20} showed in an EAGLE comparison to the Sloan Digital Sky Survey that discs generally have $\kco>0.35$ and spheroids have $\kco<0.25$.  We use the $\kco$ values calculated by \citet{davies20} using the routines of \citet{thob19}.  

\begin{table}
\caption{EAGLE galaxy kinematic morphology sample ranges for $M_\star=10^{10.7}-10^{11.2}\; \msolar$}
\begin{center}
\begin{tabular}{lccccc}
\hline
\hline
Sample & $\kco1$ & $\kco2$ & $\kco3$ & $\kco4$ & $\kco4$-med-discs\\  
$\kco$ Low & $0.114$ & $0.202$ & $0.322$ & $0.477$ & $0.477$ \\
$\kco$ High & $0.201$ & $0.321$ & $0.475$ & $0.767$ & $0.767$ \\ 
log$\langle{M}_{200}\rangle^a$ & 12.88 & 12.88 & 12.75 & 12.58 & 12.46 \\
\hline
\end{tabular}
\end{center}
\parbox{25cm}{
$^a$ Mean $M_{200}$.   
}
\label{tab:samples}
\end{table}

\begin{figure*}
  \centering
    \includegraphics[width=0.48\textwidth]{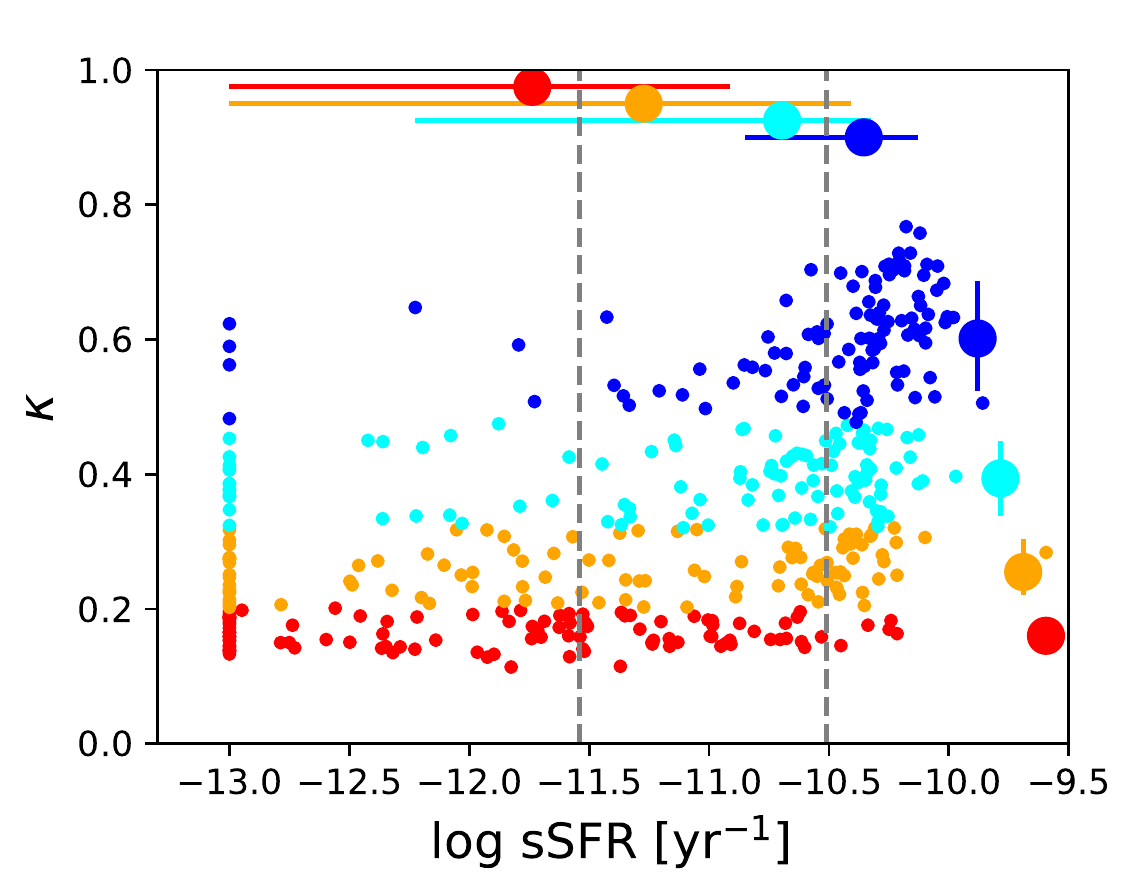}
    \includegraphics[width=0.48\textwidth]{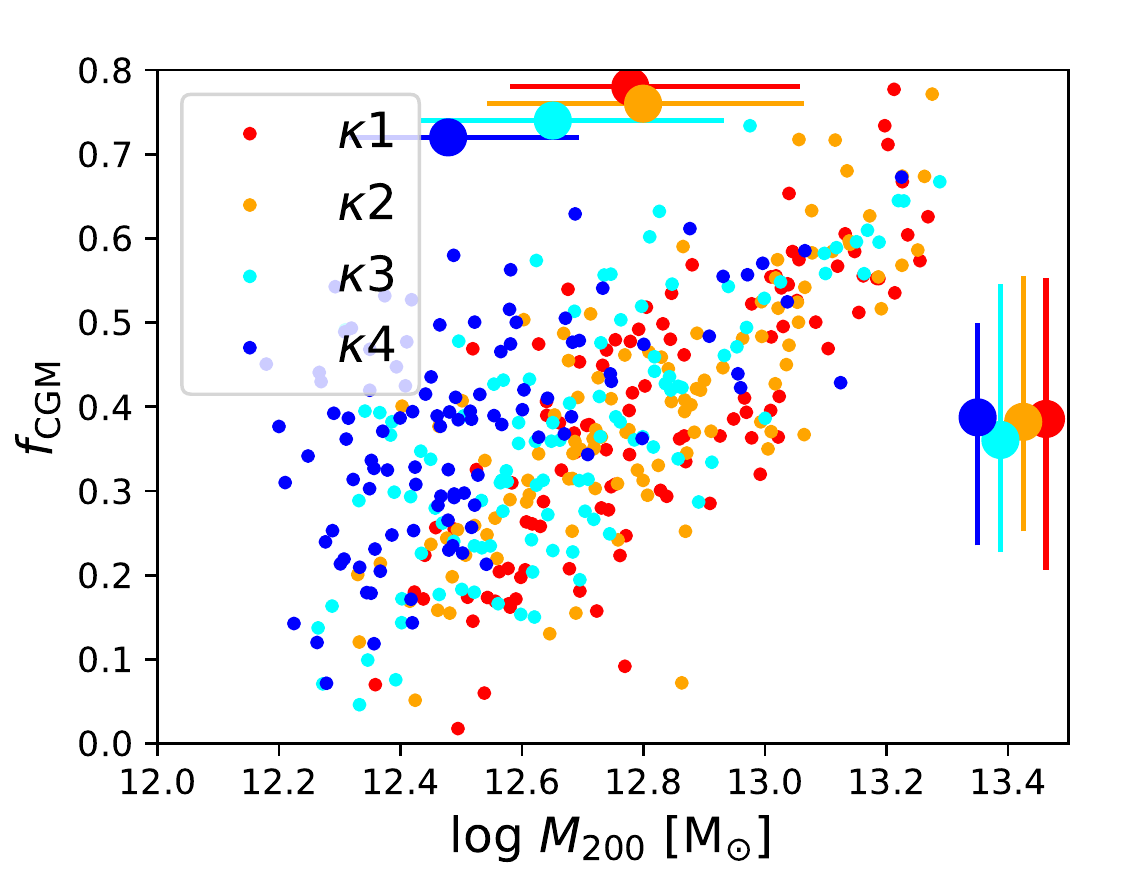}
  \caption{Our EAGLE galaxy subsamples, coloured from red to blue for $\kco1$ (spheroids) to $\kco4$ (discs).  The left panel plots two galaxy characteristics for the $M_\star=10^{10.7-11.2}\,\msolar$ samples.  sSFR correlates with $\kco$.  sSFR $\leq 10^{-13}$ yr$^{-1}$ are plotted as having $10^{-13}$ yr$^{-1}$.  The right panel plots two halo-wide characteristics, virial mass versus $f_{\rm CGM}$, which is the fractional gas content of the CGM relative to its expected total baryon content.  Large circles and error bars on the top and to the right show the median and 1-$\sigma$ dispersion for each distribution.  The dashed lines in the right panel indicate the low-sSFR (left of left line) and high-sSFR (right of right line) samples of O20.} 
  
  \label{fig:samples}
\end{figure*}

The $\kco4$ sample is clearly within the disc regime, thus we refer to this sample as ``discs.''  $\kco1$ is safely within the spheroid regime, and we refer to them as ``spheroids.''  The $\kco2$ and $\kco3$ samples are intermediate.  The left panel of Figure \ref{fig:samples} shows that $\kco$ correlates with specific star formation rate (sSFR$\equiv$SFR/$M_\star$), albeit with significant scatter.  As in O20, we select our samples based on observationally-derivable galaxy properties, which in this case is $M_\star$ and $\kco$, since as shown by \citet{thob19} $\kco$ correlates strongly with the observable ratio of rotational and dispersion velocities.  

The right panel of Fig. \ref{fig:samples} shows the gas mass fraction the of CGM, defined as 
\begin{equation} \label{equ:fCGM}
f_{\rm CGM} \equiv \frac{M_{\rm gas}(R<R_{200})}{M_{200}(R<R_{200})} \times \frac{\Omega_{\rm M}}{\Omega_{\rm b}},
\end{equation}
and the halo mass, $M_{200}$, properties that are easily definable in a simulation, but which are almost always observationally elusive. The median $M_{200}$ is higher for spheroids ($10^{12.78}\, \msolar$ for $\kco1$)  than for discs ($10^{12.48}\, \msolar$ for $\kco4$). Across the whole dataset, $f_{\rm CGM}$ increases as a function of $M_{200}$, though since each subsample spans a wide range in $M_{200}$ their median $f_{\rm CGM}$ values are similar. At fixed halo mass, discs occupy more gas-rich haloes than spheroids, a correlation shown for the whole EAGLE Ref-L100N1504 population by \citet{davies20}.

\subsubsection{Rotating galaxies}

Every galaxy is rotated edge-on using the angular momentum axis calculated from all stars within 30 kpc of its centre.  We perform this even for spheroid galaxies, but we will show there is no preferred alignment in any property we explore for the spheroids.  All images place the angular momentum axis vertically, and the disc axis horizontally.  We use a coordinate system of $\phi = 0-90^\circ$ with $0^\circ$ ($90^\circ$) being the semi-major equatorial (semi-minor polar) axis.  For disc galaxies, cosmologically-based simulations generally indicate accretion along the semi-major axis \citep{stewart17,ho20,trapp22} and superwind outflows along the semi-minor axis \citep{shen13,mitchell20a,peroux20}.

\subsubsection{The $\kco 4$-med-discs sample}

We create a separate ``medium discs'' or med-discs sample of 30 $\kco4$ galaxies human-classified based on their morphologies.  Three people, E. Huscher, A. Nica, and B. Oppenheimer, visually classified the galaxies using their total gas maps.  We i) determined if the galaxies are rotated to be well-aligned edge-on, ii) rejected galaxies with obvious gas-rich satellites, and iii) estimated the size of the discs with a score between 1 (small) and 3 (large).   We aggregated the scores, and determined that the intermediate-sized discs are most comparable to typical spiral galaxies with $\HI$ discs, such as NGC 891.  Smaller discs are compact spirals often with centralised star formation, and larger discs are often diffuse, flocculent, and/or warped, which are often in higher mass haloes.  

The human classification of the $\kco4$-med-discs sample created some surprisingly rigid data cuts.  While the median halo mass is not that much lower than the $\kco4$ sample, cf. $M_{200}=10^{12.45}$ and $10^{12.48}\,\msolar$, the mean mass is $0.12$ dex lower as listed in Table \ref{tab:samples} with a total range spanning $10^{12.20-12.75}\, \msolar$, which excises the 13 most massive $\kco4$ haloes.  The total sSFR range is $10^{-10.58}-10^{-10.00}$ yr$^{-1}$, which is a much narrower range than Fig. \ref{fig:samples} shows for $\kco4$ and excises the 29 least star-forming galaxies.

\subsection{Forward modelling pipeline}

The forward modelling pipeline we use was introduced in \S2.3 of O20.  We use the pyXSIM package\footnote{\url{http://hea-www.cfa.harvard.edu/~jzuhone/pyxsim/} pyXSIM is an implementation of the PHOX algorithm \citep{biffi12,biffi13}.} \citep{zuhone16} to create mock SIMPUT\footnote{\url{http://hea-www.harvard.edu/heasarc/formats/simput-1.1.0.pdf}} files.  For each SPH particle with $T>10^{5.3}\,$K and hydrogen number density $\nh<0.22~\cmc$ inside $3\,R_{200}$, pyXSIM generates a Monte-Carlo random sampling of photons using X-ray spectra from the Astrophysical Plasma Emission Code \citep[{\sc APEC};][]{smith01}.  {\sc APEC} assumes collisional ionization equilibrium given the density, temperature, and metallicity (including 9 individually-tracked abundances) of each SPH particle.  Like O20, we do not simulate X-rays from the ISM.  The photons from the galaxy's CGM are termed ``source'' photons.   

We place the simulated galaxies at $z=0.005$, correspond to a distance of 22.2 Mpc for our {\it eROSITA} mocks.  We include simulated Galactic foreground emission and a Cosmic X-ray background (CXB) randomly-generated using the SOXS package\footnote{\url{http://hea-www.cfa.harvard.edu/~jzuhone/soxs/}; background described in \url{http://hea-www.cfa.harvard.edu/~jzuhone/soxs/users_guide/background.html}}.  Galactic absorption assuming a column of $N_{\rm HI}=2\times10^{20}\,\cms$ is then applied.  

The SIXTE simulation software \citep{sixte} uses SIMPUT file inputs to create mock 2 kilosecond {\it eROSITA} observations with instrumental background centred on the position of the galaxy.  Event files are created using the {\tt erosim} tool for the seven {\it eROSITA} cameras and combined into one image (see fig. 1 of O20).  Individual CXB compact sources are common within each mock observation, therefore we use the CIAO \citep{ciao} {\tt wavdetect} routine to detect and mask compact sources, including CXB sources, bright satellites, and point source-like emission from dense gas at the position of the galaxy.  We mainly report on extended emission beyond a projected radius of $r\ga 10$ kpc, but plot emission profiles to 5 kpc.  We include only non-star-forming gas with density $\nh<0.22~\cmc$, because our focus is mainly on extended CGM emission as in O20.  These higher density regions correspond to locations of ISM gas as well as the stellar component that includes X-ray binary emission, both of which we also do not attempt to simulate.  Hence, our designed experiment aims to resolve CGM gas and works only at lower redshifts.   

\begin{figure}
\includegraphics[width=0.48\textwidth]{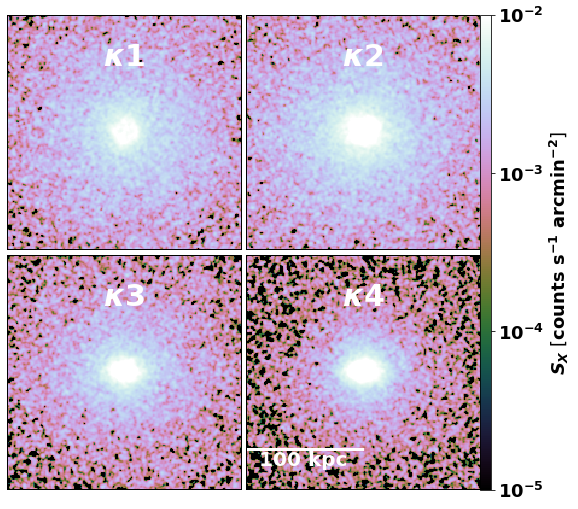}
\caption{Soft X-ray photon flux of $\kappa1$ (top left), $\kappa2$ (top right), $\kappa3$ (bottom left), and $\kappa4$ (bottom right) X-ray maps in the 0.2-1.5 keV energy band. The flux becomes more elliptical along the semi-major axis at greater $\kappa$.  
}
\label{fig:obsmaps}
\end{figure}

Individual masked images with $9.6''$ pixels are added together in our mock stacks, as are the individual exposure maps that include the {\tt wavdetect}-generated masks.  We make an off-source ``bkgd'' stack using the same procedure performed without CGM emission.  Both stacks with photon counts are divided by their respective summed exposure (``expo'') maps (in seconds) to obtain a ``signal'' photons$\,{\rm s}^{-1}$ map, using ${\rm photons}_{\rm source}/{\rm expo}_{\rm source}-{\rm photons}_{\rm bkgd}/{\rm expo}_{\rm bkgd}$.  We convert to $\photonssarcmin$ as our primary unit.  The four signal maps of 107 galaxy stacks are shown in Figure \ref{fig:obsmaps} with the disc plane aligned horizontally. 

\subsubsection{Frequency of edge-on galaxies in the observed sky}

Although we place the simulated galaxies at 22.2 Mpc, this is far too nearby to find $10^2$ edge-on galaxies.  O20 placed galaxies at $z=0.01$ with random orientations, noting that one expects there to be 70 galaxies for our mass range at an average distance of $44.6$ Mpc ($z=0.01$) across the entire sky with galactic latitude $\mid b\mid >15^\circ$.  Unlike O20, our strategy does not yield a wholly realistic observational sample.  Being an average of $2\times$ closer at $z=0.005$, this reduces the amount of galaxies by $8\times$ to 10 total galaxies.  If we assume that galaxies must have inclinations above $i=75^\circ$ to be considered edge-on, which corresponds to 26\% of galaxies, this leaves an expectations of just above 2 galaxies in our mass range at a volume-weighted average redshift of $z=0.005$.  Hence, there exists nearly a $50\times$ difference between our simulated and a realistic sample in the eRASS:8 survey.  

There are a couple ways to approach our simulated sample.  We could motivate the need to observe several nearby edge-on galaxies for a total of 200 kiloseconds after the 4-year eRASS:8 survey to achieve the expected throughput.  This could be an efficient experiment given {\it eROSITA}'s lower and more stable background as well as its superior grasp compared to {\it XMM-Newton}.  However, this does not produce a statistical ensemble using so few galaxies.  

For a larger sample, we can extend to higher redshift given our goal of resolving outside $r=10$ kpc.  Using {\it eROSITA} resolution of $15$ arcseconds, we can include galaxies up to a higher redshift, $z\approx 0.03$.  This allows $\sim$200$\times$ more galaxies with $i>75^\circ$, which creates an ensemble stack of $\sim$500 galaxies with resolvable CGM emission.  The trade-off is that the signal per galaxy declines by distance squared, and the signal is still $\sim$10$\times$ too low for eRASS:8.  Therefore, for the stacked results in \S\ref{sec:obsangular}, it is helpful to keep in mind that error ranges may be $3\times$ larger for a realistic sample from eRASS:8 assuming Poisson statistics.  

\section{Results} \label{sec:results}

In this Section, we begin by discussing results of the mock observations in \S\ref{sec:obsres}.  We focus on the general trends of our four subsamples of galaxies first, and then discuss the azimuthal dependence of X-ray emitting gas around aligned discs.  Within this section, we discuss global quantities of the galactic haloes, including $M_{200}$ and $f_{\rm CGM}$.  We then explore the physical characteristics within the EAGLE simulations in \S\ref{sec:physres} to understand the state of the CGM that gives rise to the results in \S\ref{sec:obsres}.  

\subsection{Mock Observational Results} \label{sec:obsres}

\begin{figure}
\includegraphics[width=0.48\textwidth]{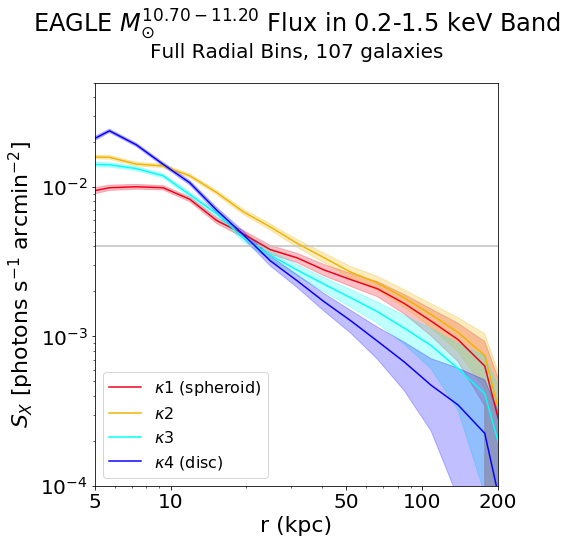}
\caption{The flux across the entire 0.2-1.5 keV band as a function of radius for the 4 subsamples with shading indicating 1-$\sigma$ Poisson errors from the source and background stacks added in quadrature. Rotationally-supported disc galaxies have greater flux at small radii than dispersion-dominated spheroid galaxies. All profiles decline in flux at larger radii, however dispersion-dominated galaxies decline more slowly.   The total astrophysical and instrumental background at 0.2-1.5 keV is indicated by grey line.}
\label{fig:obsrad_all}
\end{figure}

Figure \ref{fig:obsmaps} demonstrates the spheroid-dominated galaxies in $\kco1$ and $\kco2$ have more extended emission than the disc sample in $\kco4$.  When we plot surface brightness (SB) radial profiles in Figure \ref{fig:obsrad_all}, we see $\kco1$, $\kco3$, and $\kco4$ form a progression of declining surface brightness at large radius (beyond 20 kpc), but increasing interior emission (within 20 kpc).  The interior emission is generally more associated with the feedback-driven baryon cycle of gas outflowing, recycling, and accreting \citep[e.g.][]{mitchell20a}, while the exterior emission is more likely to arise from a quasi-static hot halo \citep[e.g.][]{opp18b}.  Discs therefore have steeper radial X-ray profiles than spheroids.  We predict the flux of spheroids at 100 kpc is $1.3\times 10^{-3}$ compared to $5\times 10^{-4}\,\photonssarcmin$ for discs, and that our experiment should be able to distinguish the two given the shaded Poisson error bar ranges generated from source plus background noise counts added in quadrature.  These levels lie below the total instrumental plus astrophysical background of $4\times 10^{-3}\,\photonssarcmin$ plotted in grey, which indicates the necessity of stable background subtraction to reveal the signal.  

The overall luminosities between 10-150 kpc for these X-ray CGMs using the 0.2-1.5 keV band are $10^{40.9}\, \lxunit$ for $\kco1$ and $10^{40.6}\, \lxunit$ for $\kco4$.  While the extended emission from spheroid CGMs averages double the luminosity of disc CGMs in the $M_\star=10^{10.7-11.2} \msolar$ bin, in \S\ref{sec:discuss_haloes} we discuss that the bias of shifted stellar mass distributions within this bin does not necessarily indicate brighter spheroid halos at a specific $M_\star$.  We also find more X-ray luminosity in the interior $10$ kpc for discs than spheroids (cf. $10^{39.6}$ and $10^{39.3} \lxunit$), but we caution that this emission will likely be drowned out by X-ray binaries, hot ISM, and potential galactic AGN in real observations.  

Finally, the $\kco2$ sample is out of sequence with brighter emission than this sequence expects.  This in part owes to this subsample having a slightly more massive average halo than the $\kco1$ subsample.  We discuss in \S\ref{sec:discuss_haloes} how $M_{200}$ and $f_{\rm CGM}$ are the best predictors for X-ray CGM luminosity. 

\subsubsection{Energy bands}

We next break down the emission by energy band, where we selected three energy bands ($0.2-0.5$, $0.5-0.8$, \& $0.8-1.5$ keV) that have approximately equal numbers of photons in our {\it eROSITA} mocks\footnote{In practice, the $0.2-0.5$ keV band is going to be highly dependent on Galactic absorption, which is assumed to be $N_{\HI}=2\times 10^{20}\,\cms$.  Owing to this absorption and {\it eROSITA}'s declining response below $<0.5$ keV, most source photons do not get counted.}.  We select these bands to optimise the division of galactic emission, which differs from the $0.3-0.6$, $0.6-1.0$, and $1.0-2.3$ keV bands used in the {\it eROSITA} all-sky map press release\footnote{https://www.mpe.mpg.de/7461761/news20200619}.

\begin{figure}
\includegraphics[width=0.48\textwidth]{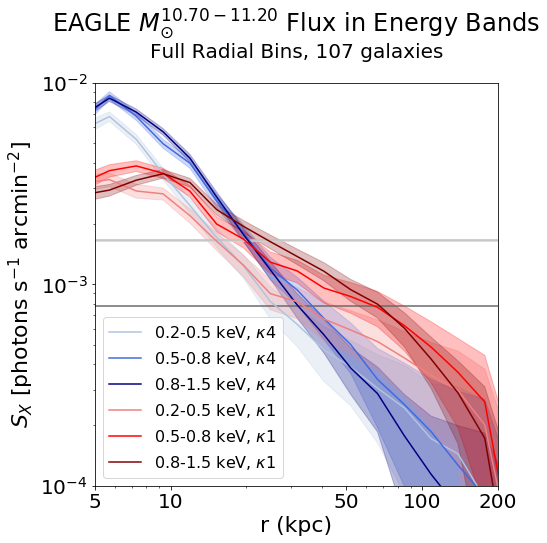}
\caption{Flux as a function of radius for rotationally-supported $\kappa4$ and dispersion-dominated $\kappa1$ galaxies divided into three energy bands. The flux in all three energy bands of $\kappa4$ is greater than $\kappa1$ at inner radii and declines at outer radii.  Background levels for individual bands are indicated by increasingly darker grey lines for higher energies (the 0.2-0.5 and 0.5-0.8 keV bands overlap). }
\label{fig:obsrad_energy}
\end{figure}

Figure \ref{fig:obsrad_energy} plots the $\kco1$ and $\kco4$ samples in red and blue respectively, with darker profiles indicating harder energy bands.  Harder X-ray emission contributes more of the surface brightness in the interior, with the two softer bands contributing relatively more in the exterior, and also in the interior of spheroids.

\begin{figure}
\begin{center}
\includegraphics[width=0.43\textwidth]{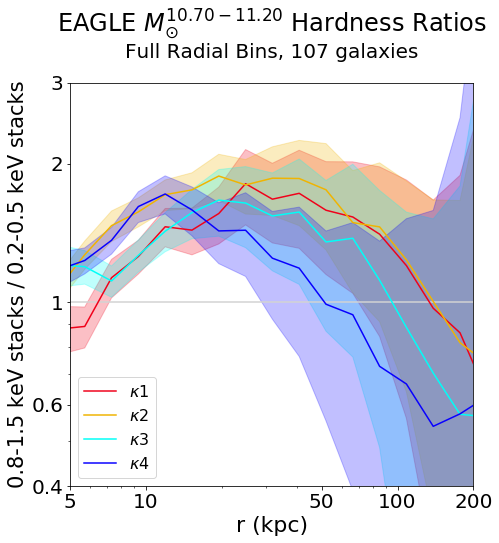}
\end{center}
\caption{Hardness ratios (ratio of highest-energy stacks to lowest-energy stacks) of the galaxy subsamples. High-$\kappa$ spiral (disc) galaxies tend to have hotter gas in the center, while low-$\kappa$ elliptical (spheroid) galaxies tend to have hotter gas in their exteriors than the spirals.  The Poisson errors here show how difficult it is to distinguish hardness ratios between samples.   }
\label{fig:obsrad_hardness}
\end{figure}

We plot the hardness ratio of the $0.8-1.5$ keV band divided by the $0.2-0.5$ keV band in Figure \ref{fig:obsrad_hardness}.  The extended CGMs of spheroids are harder, while the interior emission around discs is harder.  The overlapping 1-$\sigma$ Poisson errors of this ratio indicate this is a challenging measurement.  Furthermore, this ratio is not very sensitive to temperature, being primarily affected by metal emission that contributes disproportionally to the $0.8-1.5$ keV especially in the cooler gas around discs, but we will discuss that there are still indications of differing temperature structures in \S\ref{sec:physres}.  Achieving such a measurement requires correcting for Galactic absorption in a consistent way, as the low-energy photon counts are heavily dependent on foreground absorption.

\subsubsection{Azimuthal dependence} \label{sec:obsangular}

We now divide the galaxy into azimuthal regions to explore the orientation dependence of X-ray emission around edge-on galaxies, which is more pronounced in the disc galaxy sample.  We define two regions, the ``equatorial'' direction along the semi-major axis, and the ``polar'' direction along the semi-minor axis.  We attempt different opening angles, including $\Phi = \pi/2$, which includes all emission within the polar and equatorial directions, as well as smaller opening angles.  A smaller opening angle yields a greater difference between the equatorial and polar regions for disk galaxies that show azimuthal dependence, since smaller angles more exclusively capture the edges of the edge-on galaxies in the equatorial region and the outflows in the polar region, so the azimuthal dependences are not ``averaged out'' by the intermediate regions.  We choose to focus our results using $\Phi = \pi/4$, which includes angles between $\phi=0-22.5^\circ$ and $67.5-90^\circ$ while discarding intermediate angles.  

\begin{figure}
\includegraphics[width=0.48\textwidth]{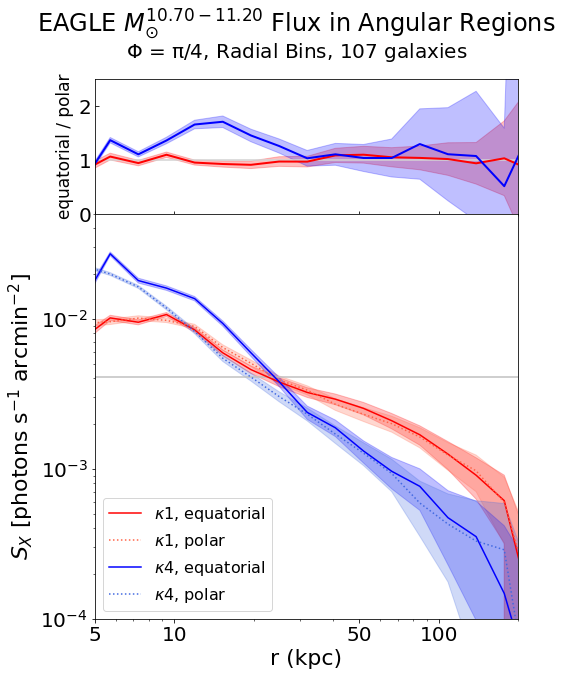}
\caption{Flux divided into equatorial and polar axis regions using $\Phi=\pi/4$ regions.  Disc galaxies ($\kco 4$) show greater flux along equatorial than polar regions from $10-30$ kpc.  Spheroidal galaxies ($\kco 1$) unsurprisingly do not show any azimuthal dependence, even though they are rotated according to their stellar kinematics.  The top panel shows the ratio of the equatorial to polar emission.   
}
\label{fig:obsrad_angle}
\end{figure}

Figure \ref{fig:obsrad_angle} shows the equatorial and polar surface brightness radial profiles for the spheroids and discs.  While we do not see any dependence for the spheroids, as expected, we recover a clear azimuthal dependence for the discs.  The equatorial axis is brighter between 5-30 kpc, as much as 60\% ($0.2$ dex) at $r\approx 15$ kpc, as indicated by the ratio in the top panel.  We also note that the elongation is visible in the Fig. \ref{fig:obsmaps} $\kco4$ stack. 

\begin{figure}
\includegraphics[width=0.48\textwidth]{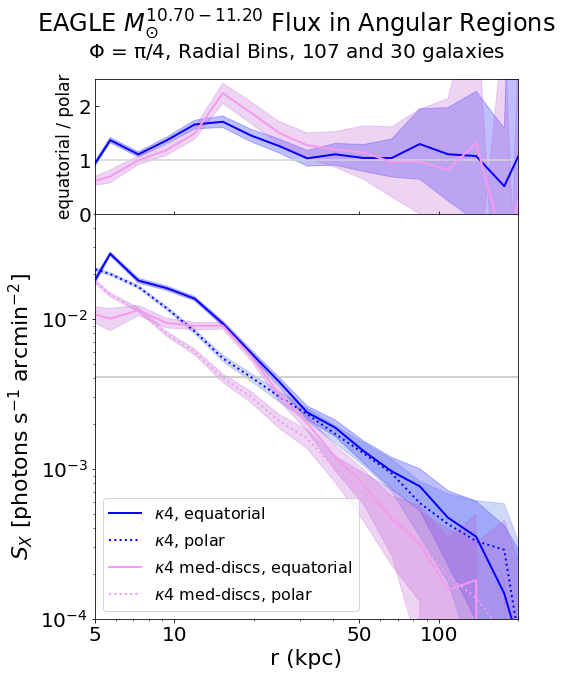}
\caption{The 30 galaxy med-discs sample, selected to be similar to some of the most well-observed edge-on discs from the $\kco4$ sample, is shown in purple.  Longer exposures are used to match the same total exposure time of the $\kco4$ sample.  The azimuthal dependence is greater that the $\kco4$ sample, with the equatorial flux being double the polar flux at $\sim 15$ kpc. 
}
\label{fig:obsrad_angle_meddisks}
\end{figure}

In Figure \ref{fig:obsrad_angle_meddisks} we show that the 30 med-discs sample exhibits greater azimuthal surface brightness dependence, although weaker overall luminosity, because this sample is preferentially devoid of the more massive haloes of the $\kco 4$ sample; the mean halo mass is $M_{200}=10^{12.46}\, \msolar$ versus $10^{12.58}\, \msolar$ for the entire $\kco4$ sample.  We selected the med-discs sample for a cleaner sample of edge-on galaxies, and we find that the azimuthal dependence is greater, with more emission along the disc axis.  

\begin{figure*}
\includegraphics[width=0.32\textwidth]{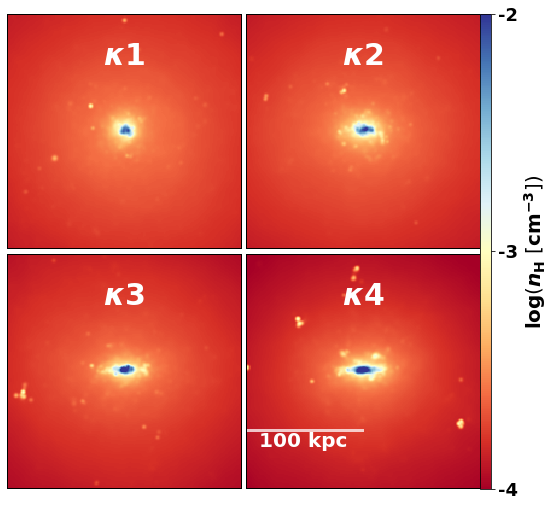}
\includegraphics[width=0.325\textwidth]{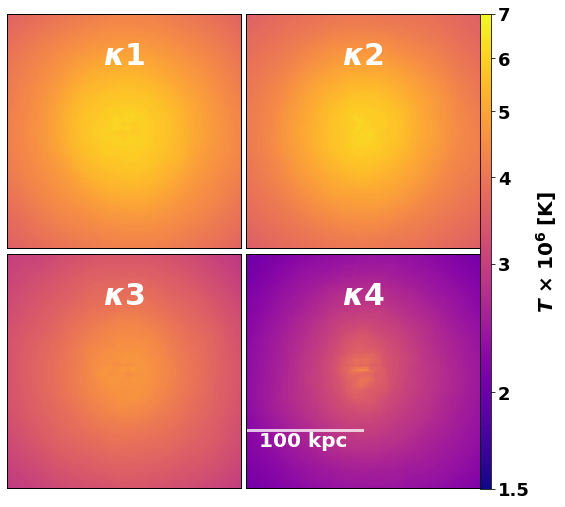}
\includegraphics[width=0.335\textwidth]{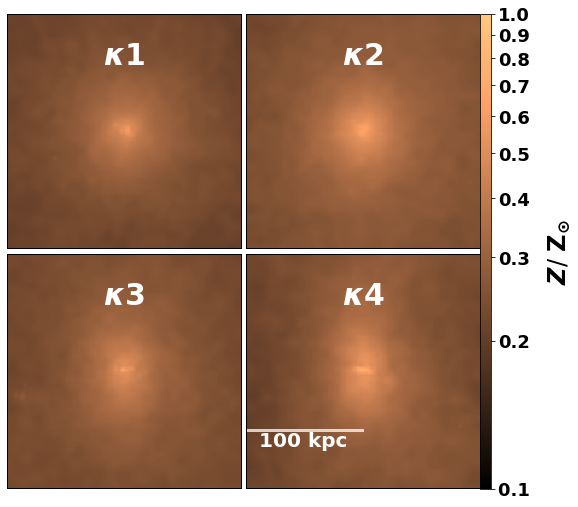}
\caption{Physical property stacked maps of the hot gas within $200\times 200$ kpc panels: hydrogen number density (left panels), temperature (middle panels), and metallicity (right panels).   Each panel set shows the 4 subsamples: $\kappa1$ (top left), $\kappa2$ (top right), $\kappa3$ (bottom left), and $\kappa4$ (bottom right).  Higher $\kappa$ galaxies have more elliptical extended gas CGMs beyond the disc, which are slightly lower density than spheroids.  The temperature panels show cooler gas around discs than spheroids, indicative of discs living in lower mass haloes.  Bipolar metal-enriched outflows are apparent in disc galaxies.}
\label{fig:physicalmaps}
\end{figure*}

The brighter equatorial emission beyond the optical disc of the galaxy may be surprising if one expects more X-rays from bipolar outflows.  In fact, \citet{truong21} does find brighter polar emission at greater radii in IllustrisTNG and EAGLE, which we discuss in \S\ref{sec:truong2021}.  Observationally, bipolar outflows have been observed around starbursting galaxies, extending out to 5-10 kpc as in the case of M82 and NGC 253 and others \citep{strickland04}, and even out to 30 kpc in the recent results of NGC 3079 \citep{hodgeskluck20}.  More typical galaxies do not as often exhibit extended bipolar emission, although bipolar outflows are sometimes observed on smaller scales as in M31 \citet{bogdan08}.  Our prediction of equatorially-enhanced emission occurs at a larger scale ($10-30$ kpc), which is mainly below the detection threshold of X-ray CGMs around other edge-on galaxies, although deep {\it XMM-Newton} observations by \citet{hodgeskluck18} for NGC 891 may provide upper limit constraints, already, as we discuss in \S\ref{sec:discuss_lum}.  

Finally, we have explored hardness ratios in azimuthal regions, but did not detect any observationally detectable variation.  This may reflect the weak azimuthal temperature dependence as we will explore in the next subsection.

\subsection{Physical Properties } \label{sec:physres}

We now discuss the underlying physical properties of the hot gas that give rise to the X-ray emission around the stacked galaxies.  We begin by looking at averaged maps and then move to radial profiles.  

\subsubsection{Physical property maps}

The gas density maps in Figure \ref{fig:physicalmaps} (left panels) include gas only above $T=10^{5.3}$ K (the cutoff temperature of the {\sc APEC} tables).  In addition to subtle changes, these maps show an obvious disc-like structure in the $\kco4$ map, which has slightly lower extended gas density as well.  The extended $\kco4$ hot CGM has some ellipticity along the disc axis.  The temperature maps (middle panels) show far more variation, and the lower temperatures around disc galaxies are a sign of their lower halo masses in Fig. \ref{fig:samples}. There exists little azimuthal dependence in temperature beyond the disc.  The metallicity maps, again of only the $T\geq 10^{5.3}$ K gas and normalised to solar using \citet{asplund09} abundances, in the right panels show a polar enhancement indicating enrichment by bipolar outflows for the $\kco4$ sample.  \citet{crain13} stacked GIMIC galaxies in much the same way, but did not see bipolar metal outflows in these simulations with only stellar feedback that was weaker than EAGLE's prescription.  We note similar trends in EAGLE as seen in IllustrisTNG by \citet{truong21}, which we discuss further in \S\ref{sec:truong2021}.  

\subsubsection{Azimuthal radial profiles}

\begin{figure*}
\includegraphics[width=0.34\textwidth]{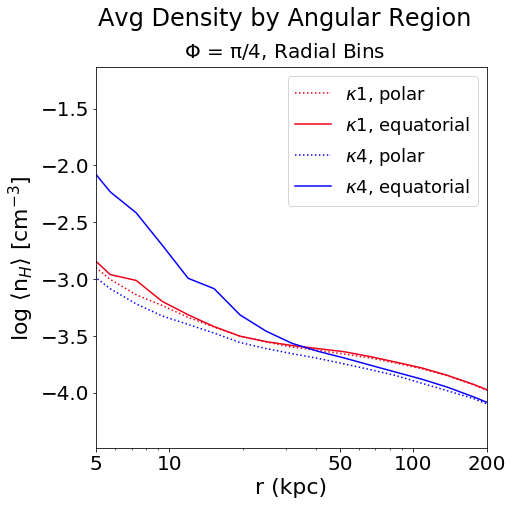}
\includegraphics[width=0.32\textwidth]{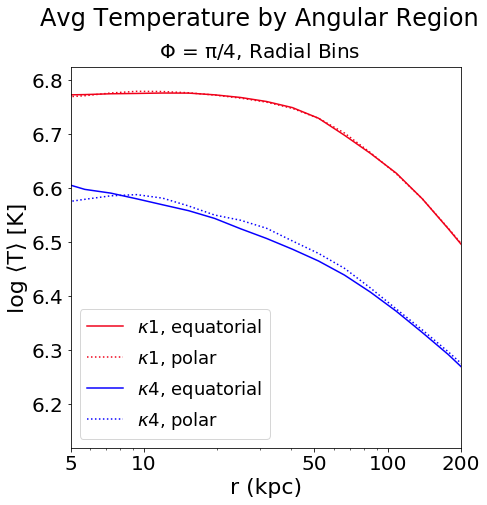}
\includegraphics[width=0.32\textwidth]{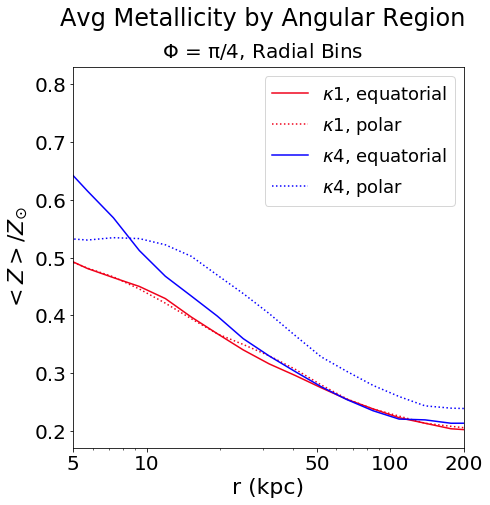}
\caption{Projected radial profiles in polar and equatorial azimuthal bins of physical quantities for the $\kco1$ and $\kco4$ samples.  Mean density, temperature, and metallicity are displayed from left to right.  $\kco1$ stacks show no azimuthal dependence as expected, while the $\kco4$ stacks show several differences, the most significant being density.  Metallicity, normalised using $\Zsolar=0.142$, is enhanced along the polar direction indicating outflowing gas.  }
\label{fig:physical_ang}
\end{figure*}

We now divide the physical property maps into $\Phi=\pi/4$ regions, as we did for the flux maps in \S\ref{sec:obsangular}, and plot equatorial and polar regions for the $\kco1$ and $\kco4$ stacks in Figure \ref{fig:physical_ang}.  Only $\kco4$ shows azimuthal dependence, as expected.  The density (right panel) indicates the most significant difference with $2.4\times$ ($1.6\times$) higher density along the equatorial direction than the polar direction at 15 (20) kpc.  The temperatures do not show significant azimuthal dependence, while the metallicity is enhanced along the bipolar outflow direction.  

It is interesting that the greatest difference is in density, and if X-ray emission is scaled by density squared, one may expect an even larger difference than the 60\% difference at 15 kpc.  The slightly lower metallicity is not enough to explain the difference.  Firstly, the distribution of physical properties (i.e. individual gas particles in the simulation) determines X-ray emission, and not the average.  Additionally, we apply the {\tt wavdetect} algorithm to excise point-like sources, even if they arise from dense concentrations of CGM gas, which could boost average density but not contribute to X-ray emission.

\section{Discussion} \label{sec:discuss}  

Our main results are that 1) the X-ray CGM of spheroidal galaxies in a fixed stellar mass bin appear more luminous than that of discs, and 2) edge-on discs exhibit azimuthal dependence where X-ray emission is brighter along the equatorial axis than the polar direction.  

\subsection{Why spheroids appear to have brighter X-ray CGMs than discs} \label{sec:discuss_haloes}

O20 defined the extended X-ray emission as beyond 10 kpc, $\LXextended$, finding that high-sSFR CGMs are brighter than their low-sSFR counterparts, although the difference is small ($\la 0.1$ dex) for the O20 EAGLE high-mass samples.  However, our morphological samples show that spheroids have an average $\LXextended=10^{40.9}\,\lxunit$ that is double the brightness of discs with $\LXextended=10^{40.6}\,\lxunit$.\footnote{We integrate luminosities between 10-150 kpc using the $0.2-1.5$ keV band, which is shifted from the O20 using 10-200 kpc and the $0.5-2.0$ keV band, but are of comparable luminosity. }   Given that discs (spheroids) are generally star-forming (passive), this duality in average extended luminosities presents a paradox-- how do both high-sSFR and spheroids possess brighter X-ray CGMs?  

O20 performed linear regressions on the halo properties that are most predictive for X-ray luminosity (their fig. 4), and found that a relation where
\begin{equation} \label{equ:LX}
\LXextended=L_{X,0} M_{200}^\alpha f_{\rm CGM}^\beta
\end{equation}
well-describes extended X-ray CGM emission.  In this formulation, $\alpha$ ranged between $1.2-1.6$ for EAGLE and IllustrisTNG haloes hosting $M_\star\approx 10^{10.2-11.2}\, \msolar$  galaxies with $\alpha=1.2$ being the relation for the EAGLE sample we explore here, and $\beta$ ranged between $1.6-2.0$ with $\beta= 2.0$ for our sample.  Hence, the total gas fraction inside $R_{200}$ is a greater determinant than halo mass, especially for the EAGLE high-mass sample.  As discussed in \S\ref{sec:samples} using Fig. \ref{fig:samples}, both the median and mean $M_{200}$ is $0.3$ dex higher for spheroids while median $f_{\rm CGM}$ is similar; therefore it is consistent with Equation \ref{equ:LX} that the spheroids having $\sim 0.3$ dex brighter luminosities owes mainly to halo mass.  Compare this to the O20 sSFR division, where the high-sSFR bin has a median $f_{\rm CGM}$ that is $0.13$ dex higher and $M_{200}$ that is $0.10$ dex lower versus the low-sSFR sample; therefore the high-sSFR sample is just slightly ($\sim 0.1$ dex) brighter according to Equ. \ref{equ:LX} that supports the O20 result.  

\begin{figure}
    \centering
      \includegraphics[width=0.48\textwidth]{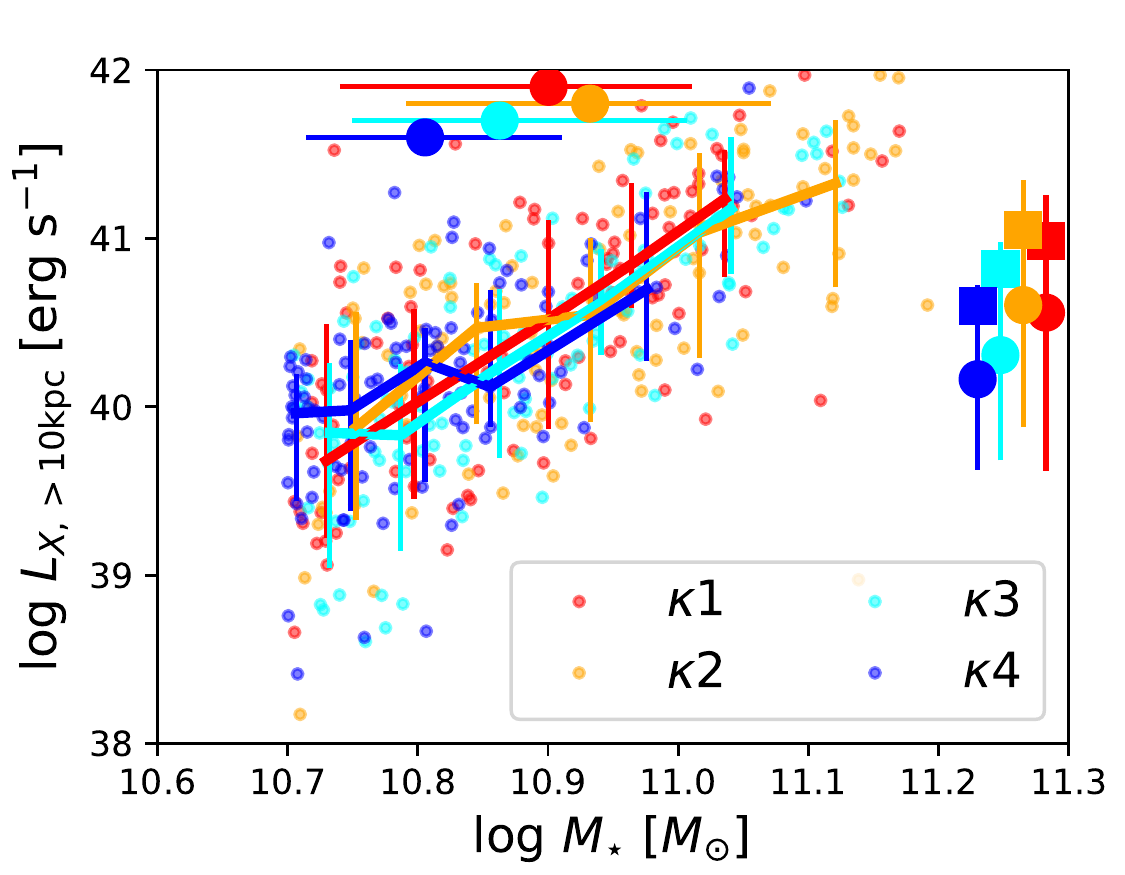}
    \caption{Extended ($>10$ kpc) X-ray emission as a function of stellar mass divided into our 4 morphological samples with points showing individual galaxies and solid lines showing medians.  Lines are adaptive bins that divide the 107-galaxy samples into 5 equally sized $M_\star$ bins with vertical error bars showing the 1-$\sigma$ spread.  There is no obvious trend for $\LXextended$ on morphology at fixed $M_\star$.  Large circles and error bars on the top and to the right show the median and 1-$\sigma$ dispersion for each distribution.  Brighter spheroid haloes have a higher distribution of $M_\star$ than disc haloes.  Large squares indicate the mean $\LXextended$.}
    
    \label{fig:mstar_LX}
\end{figure}

While the spheroids appear brighter than discs owing to higher halo masses, even within fixed $M_\star$ bins that span a factor of three in mass, the distributions of stellar masses are not uniform.  We plot individual $\LXextended$ values calculated directly from the SIMPUT files\footnote{As in O20 fig. 4, we take luminosities calculated by pyXSIM before they are put through the SIXTE instrument simulator.} as a function of $M_\star$ in Figure \ref{fig:mstar_LX}.  We find no obvious trend of luminosity with morphology at fixed $M_\star$.  Instead, the distribution of stellar masses within the bin is the most important determinant for the average $\kco$ sample luminosity, with $\kco2$ ($\kco4$) having the most (least) massive galaxies.  Therefore, we find that spheroids {\it at fixed specific $M_\star$} are not necessarily brighter than discs in EAGLE.  Higher stellar masses are indicative of higher halo masses for our stellar mass-defined samples.  

As a consistency check, the squares plotted in Fig. \ref{fig:mstar_LX} are the mean $\LXextended$ values for each morphological sample calculated from the SIMPUT files.  They agree with the $\LXextended$ values from the forward-modeled surface brightness profiles in Fig. \ref{fig:obsrad_all} within 0.1 dex, demonstrating that our stacking method recovers the true answer from the simulation.

This exercise using a simulation where we know the answers provides a cautionary tale when interpreting observations.  The spheroid CGMs, which appear double the brightness as their disc counterparts, are not necessarily brighter at a specific $M_\star$.  This demonstrates that it is possible to reach a false conclusion through stacking; however one can sub-divide their samples many different ways (e.g. different stellar mass bins in this case) to test the robustness of their conclusions when working with real observations where one does not have the answer from individual galaxies.

Related to this point, one might also get the impression that $f_{\rm CGM}$ is insensitive to morphology from the nearly identical median $f_{\rm CGM}$ for the four samples. However, as shown in the right panel of Fig. \ref{fig:samples} \citep[and by][]{davies20}, spheroids in fact have lower $f_{\rm CGM}$ than discs at fixed halo mass. The similar median values across the samples occur because spheroids have below-average $f_{\rm CGM}$ values at higher $M_{200}$, while discs have above-average $f_{\rm CGM}$ values at lower $M_{200}$. At fixed halo mass, the CGM around discs should be \textit{brighter} than that around spheroids, however in our kinematically-defined samples the effect of halo mass causes spheroids to appear brighter. 

\subsection{Equatorially-enhanced X-ray emission}

Our prediction that extended diffuse X-ray haloes at $10-30$ kpc should be brighter along the semi-major axis of their host galaxies provides a key test for future {\it eROSITA} observations.  The prediction is related to the dynamical state of hot gaseous haloes that deviate from spherical hydrostatic equilibrium by having net rotation along the disc axis \citep{opp18b}.  Given the observation that the Milky Way gaseous hot halo may be rotating \citep{hodges16}, a set of analytical models for rotating hot haloes with enhanced densities along the disc axis was developed by \citet{sormani18}.  

\begin{figure*}
\includegraphics[width=0.98\textwidth]{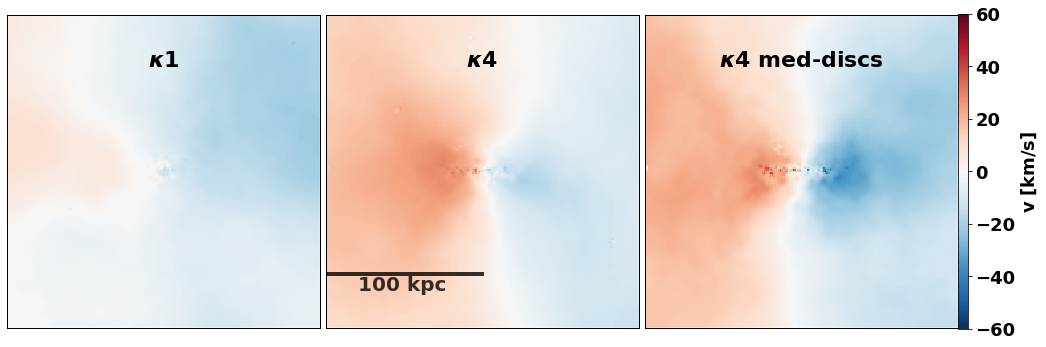}
\caption{Projected line-of-sight velocity of gas with $T>10^{5.3}$ K for the $\kco1$, $\kco4$, and $\kco4$-med-discs samples, which are rotated to have their velocities aligned.  The med-discs sample has the highest hot gas rotation aligned with the stellar disc.  The low level of preferred rotation in the $\kco1$ sample probably arises because these galaxies were also rotated based on their stellar velocities. 
}
\label{fig:velocity_maps}
\end{figure*}

We show the velocity maps for the the $\kco1$, $\kco4$, and $\kco4$-med-discs samples in Figure \ref{fig:velocity_maps}.  Indeed, there is net rotational velocity in these galaxies that were rotated and stacked to have their angular momentum vectors aligned.  While X-ray emission probes will likely not be able to observe these velocities in the foreseeable future, the determination via azimuthal emission of denser equatorial gas has important implications for how disc galaxies accrete material from the hot CGM.  If we consider the precipitation criterion of $t_{\rm cool}/t_{\rm ff} \la 10$ \citep{sharma12} for gas to cool, the rotating models of \citet{sormani18} favor condensation of cool gas near the disc axis by i) lowering the cooling time ($t_{\rm cool}$) with increased density, and ii) raising the effective free-fall time ($t_{\rm ff}$) via rotational support \citep{sormani19}.  This later paper argues cooling from the hot CGM within $\sim 30^\circ$ of the disc promotes the formation of high-velocity cloud structures.  \citet{sormani18} models predict hotter gas in the polar direction without the presence of outflows, which we do not see in Fig. \ref{fig:physical_ang}.  

Observations of inclined galaxies, including NGC 891, NGC 3079, NGC 4631, NGC 5775, NGC 5907, show enhanced emission along the disc axis, but on the scale of the optical extent of the galaxy within 10 kpc of the center \citep{li08,hodgeskluck18, hodgeskluck20}.  None of these observations are able to detect significant emission beyond 10 kpc in the equatorial direction, which may be a result of their lower halo masses than our more massive halo stack here.  \citet{juranov20} observed 6 lenticular, S0 galaxies finding enhanced equatorial emission in these galaxies, but mostly within their optical extents.  These lenticular galaxies may correspond to our $\kco2$ or even $\kco3$ galaxies, and live in halo masses at the upper range or above our simulated galaxy halo sample.  In all these cases, the X-ray emission is more associated with the hot ISM or disc-halo interface, rather than the CGM.  

One may expect polar emission from outflows, especially given that EAGLE launches thermal winds associated with star formation with $T\sim 10^{7.5}$ K \citep{mitchell20a}.  Observationally, there exists extended polar emission associated with starbursts \citep{strickland02,strickland04, strickland09,hodgeskluck20}, and even  moderately star-forming galaxies \citep{hodgeskluck18}.  However, it may well be that much of this gas expected to reside at $\sim 10^7$ K is too tenuous to be X-ray bright, therefore the bulk of the mass in the outflow remains undetected \citep{strickland00}.  Outflows along the polar direction are not detected in EAGLE in $\MgII$ absorption either \citep{ho20}, and appear to be detected in observations \citep{bordoloi11,bouche12,lan14,lan18,huang21}, which suggests that EAGLE superwind feedback may not reproduce significant aspects of observed outflows.

\subsection{Are EAGLE galactic X-ray CGMs too bright?} \label{sec:discuss_lum}

The EAGLE simulations were not calibrated to reproduce the X-ray luminosities and/or gas fractions in the group and cluster regimes \citep{crain15}, which \citet{schaye15} demonstrated produced X-ray emission that is too luminous.  

It remains less clear if CGM X-ray emission is also too luminous.  As \citet{davies19} and \citet{kelly21} showed, the \citet{anderson15} {\it ROSAT} stacking could not necessarily rule out EAGLE X-ray luminosities.

It remains less clear if X-ray CGM emission at $M_{200}\la 10^{13}\,\msolar$ are also too luminous, as \citet{davies19} and \citet{kelly21} showed, the \citet{anderson15} {\it ROSAT} stacking could not necessarily rule out EAGLE X-ray luminosities.

The best constraints for extended X-ray emission around disc galaxies are from individual targeted galaxies with {\it Chandra} and {\it XMM-Newton}, with the latter having the soft X-ray response more closely matching {\it eROSITA}.   

Only the most massive spiral galaxies, NGC 1961 \citep{anderson16}, NGC 6753 \citep{bogdan17}, NGC 3221 \citep{das19}, and the CGM-MASS sample \citep{li17} have detectable extended X-ray emission, which O20 argued are weaker than their high-sSFR {\it eROSITA} mock stacks containing similar mass galaxies.  This potential mismatch is not exclusive to EAGLE, as O20 found similar values $\LXextended$ in IllustrisTNG, which was also explored by \citet{truong19} who concentrates on centralised soft X-ray emission from IllustrisTNG galaxies.   The lack of extended emission in deep NGC 891 {\it XMM-Newton} observations \citep{hodgeskluck18} may already set an upper limit for detectable emission, which may also suggest that X-ray emisssion around discs is in reality fainter than EAGLE predicts.  It is likely, however, that NGC 891 lives in a halo at the low-mass end of the $\kco4$ stack based on its stellar mass.  

{\it eROSITA} will provide a uniform survey in which to stack many edge-on galaxies, including those at larger distances than our idealised sample at 22 Mpc that are perfectly edge-on.  In reality {\it eROSITA} will enable the stacking of galaxies at a variety of distances with high inclination angles, which should work for our proposed experiment to test the presence of ellipsoidal hot gaseous haloes.  

\subsection{Comparison with Truong et al. (2021)} \label{sec:truong2021} 

In a publication addressing similar topics, \citet[][hereafter T21]{truong21} presented physical properties and X-ray predictions of galaxies oriented to be edge-on, focusing on results from the IllustrisTNG (110 Mpc)$^3$ volume, and additional results from EAGLE.  This paper is similar in approach to our work here, but their emphasis is on the extended hot CGM between 0.25 and 0.75 $R_{200}$.  In fact, they find up to a factor $2\times$ stronger emission along the polar axis of both IllustrisTNG and EAGLE, but at $0.5 R_{200}$ around $M_{200}=10^{12.0-12.5}\,\msolar$ haloes.  While it may seem this contradicts our finding, our focus centres on equatorial enhancement at much smaller radii ($10-30$ vs. $100-150$ kpc) and at somewhat higher halo masses ($10^{12.3-12.7}$ vs. $10^{12.0-12.5}\;\msolar$) in the $\kco 4$ sample.  In fact, T21 finds very slightly enhanced equatorial emission at $0.5 R_{200}$ in EAGLE in their fig. 7 for our halo mass range.  

Nonetheless, T21 shows similar trends in the physical properties in both IllustrisTNG and EAGLE as we show in Figs. \ref{fig:physicalmaps} and \ref{fig:physical_ang}, including increased density along the equatorial axis, and increased metallicity and temperature along the polar axis.  We do note that the anisotropies in EAGLE are weaker than in IllustrisTNG based on their figs. 6 and A2.  They also divide their emission maps into bands, finding harder emission along the polar extent, which our proposed experiment would not have the signal to observe.  T21 emphasised how different X-ray signatures from IllustrisTNG, EAGLE, and even Illustris \citep{vogelsberger14} could help ascertain the nature of SMBH feedback.  For example, \citet{pillepich21} demonstrated polar signature associated with feedback events in the higher resolution IllustrisTNG50 volume, and argued that the {\it eROSITA} bubbles observed above and below the Milky Way disc \citep{predehl20} may be common around other disc galaxies and indicative of episodic jet-driven AGN events.  

We note that the T21 predictions of polar-enhanced emission are at lower masses and larger radii, corresponding to surface brightness limits far below what our proposed experiment can detect, $\la 10^{34}$ erg s$^{-1}$ kpc$^{-2}$ (see T21 maps in their fig. 7).  Hence, theirs is a different experiment stacking $\sim 10^4$ galaxies out to $z\sim 0.1$, and based on this work and O20, it may be hard to achieve this detection threshold with standard stacking techniques.  It may well be that our proposed experiment detects equatorial enhancement at lower CGM radii, while the T21 polar enhancement exists at larger radii.  

\section{Summary} \label{sec:summary}  

We apply an X-ray emission forward modeling pipeline to EAGLE galaxy haloes sorted by stellar morphology to determine how X-ray haloes depend on morphology and to assess the feasibility of detecting azimuthal dependence around edge-on spirals.  By stacking mock {\it eROSITA} observations of gaseous haloes hosting $M_\star=10^{10.7-11.2}\;\msolar$ galaxies, we predict the following:  

\begin{itemize}

\item Stacked spheroids have more extended and brighter CGMs than disc galaxies.  This owes to spheroids living in more massive haloes than discs for our fixed $M_\star$ bin.  Related, spheroids have higher average $M_\star$ within this $\Delta$log$M_\star= 0.5$ bin, but spheroids are not generally brighter than discs at the exact same $M_\star$.  On the other hand, disc galaxies with more overall star formation have brighter emission from their inner 20 kpc. (Figs. \ref{fig:obsmaps}, \ref{fig:obsrad_all}, \ref{fig:mstar_LX})

\item Edge-on disc galaxies show clear azimuthal dependence with the semi-major axis being up to $60-100\%$ brighter at $15$ kpc than the semi-minor axis.  Even though thermally-driven outflows preferentially travel perpendicular to the disc in EAGLE, this does not translate to higher X-ray emission.  The equatorial enhancement of emission appears primarily driven by greater gas densities for $T \ga 10^6$ K gas. (Figs. \ref{fig:obsrad_angle}, \ref{fig:physicalmaps}, \ref{fig:physical_ang})

\item The hotter temperatures of spheroid versus disc CGMs may be observable by taking a ratio at the high and low end of the {\it eROSITA} soft X-ray response; however, the dependence is weak and may be difficult to detect.  (Figs. \ref{fig:obsrad_energy}, \ref{fig:obsrad_hardness})

\item The denser gas along the disc axis relates to net co-rotation of these hot haloes with the stellar disc.  Although these velocities would be difficult to observe directly, we argue that brighter X-ray emission along the semi-major axis correlates with the gas rotation. (Fig. \ref{fig:velocity_maps})   

\item Our human classification technique to identify edge-on EAGLE disc galaxies creates a cleaner sample in lower mass haloes to compare to some of the best observed nearby edge-on galaxies, which we predict should have double the emission along the semi-major axis compared to the semi-minor axis at 15 kpc.  (Fig. \ref{fig:obsrad_angle_meddisks})

\item We note the results by \citet{truong21} who find brighter X-ray emission along the semi-minor axis mainly at $>$100 kpc around IllustrisTNG disc-like galaxies owing to superwind outflows.  We find similar although weaker physical property anisotropies in EAGLE as them, but we argue that this more distant polar gas is too tenuous to measure in our designed experiment.  

\end{itemize}

The detection of the extended hot CGM around inclined galaxies will provide key insights for how galaxies get their gas \citep[e.g.][]{keres05}.  While modeling hot gaseous haloes under the assumption of spherical hydrostatic equilibrium may be a good assumption for spheroid-hosting haloes, the deviation from sphericity around disc galaxies \citep{opp18b} can be tested by the axial ratios of hot haloes beyond $10$ kpc around discs.  Denser, co-rotating gas along the semi-major axis can better facilitate the cooling of the hot CGM and the condensation of $\sim 10^{4}$ K phase \citep{sormani19}.  Targeting nearby disc galaxy CGMs with {\it eROSITA} after the completion of the eRASS:8 survey may be worth the investment to ascribe a comprehensive theoretical explanation to multi-wavelength observations of the CGM.

\section*{acknowledgements}

The authors wish to acknowledge Ezra Huscher, Edmund Hodges-Kluck, Joop Schaye, Nhut Truong, and Nastasha Wijers for essential contributions to this work.  AN was supported by the University of Colorado Boulder's Undergraduate Research Opportunities Program (UROP), which provides grants to support student-faculty partnerships and projects in all fields of study.  BDO, AB, WRF, and RPK acknowledge support from the Smithsonian Institution.  RAC is a Royal Society University Research Fellow.  AB, RPK, and WRF acknowledge support from the High Resolution Camera program, part of the Chandra X-ray Observatory Center, which is operated by the Smithsonian Astrophysical Observatory for and on behalf of the National Aeronautics Space Administration under contract NAS8-03060.  JJD is supported by the European Union’s Horizon 2020 research and innovation programme under grant agreement No. 818085 GMGalaxies”.  The study used high performance computing facilities at Liverpool John Moores University, partly funded by the Royal Society and LJMU’s Faculty of Engineering and Technology.  

\section*{data availability}

All simulated mock observations and other data outputs from this work are available to the public upon request.  Please e-mail anna.nica@colorado.edu e-mail if you require any data in the figures in this paper, and benjamin.oppenheimer@colorado.edu if you require simulation require raw mock observational data of stacks or even of individual galaxies.

\bibliography{main}{}
\bibliographystyle{mnras}

\end{document}